\documentclass[twocolumn,showpacs,preprintnumbers,amsmath,amssymb,superscriptaddress]{revtex4}

\usepackage{graphicx}% Include figure files
\usepackage{dcolumn}% Align table columns on decimal point
\usepackage{bm}% bold math
\usepackage{color}

\begin{document}

\title{Spatially uniform calibration of a liquid xenon detector at low energies using $^{83\mathrm{m}}$Kr}

\author{A.~Manalaysay}
\email[Corresponding author, \\ \emph{Electronic address}: ]{aaronm@physik.uzh.ch}
\affiliation{Physik-Institut, Universit\"{a}t Z\"{u}rich, Z\"{u}rich, 8057, Switzerland}
\affiliation{Department of Physics, University of Florida, Gainesville, FL 32611, USA}

\author{T.~Marrod\'an Undagoitia}
\affiliation{Physik-Institut, Universit\"{a}t Z\"{u}rich, Z\"{u}rich, 8057, Switzerland}

\author{A.~Askin}
\affiliation{Physik-Institut, Universit\"{a}t Z\"{u}rich, Z\"{u}rich, 8057, Switzerland}

\author{L.~Baudis}
\affiliation{Physik-Institut, Universit\"{a}t Z\"{u}rich, Z\"{u}rich, 8057, Switzerland}

\author{A.~Behrens}
\affiliation{Physik-Institut, Universit\"{a}t Z\"{u}rich, Z\"{u}rich, 8057, Switzerland}

\author{A.~D.~Ferella}
\affiliation{Physik-Institut, Universit\"{a}t Z\"{u}rich, Z\"{u}rich, 8057, Switzerland}

\author{A.~Kish}
\affiliation{Physik-Institut, Universit\"{a}t Z\"{u}rich, Z\"{u}rich, 8057, Switzerland}

\author{O.~Lebeda}
\affiliation{Nuclear Physics Institute, Academy of Sciences of the Czech Republic, CZ 250 68, \v{R}e\v{z} near Prague, Czech Republic}

\author{R.~Santorelli}
\affiliation{Physik-Institut, Universit\"{a}t Z\"{u}rich, Z\"{u}rich, 8057, Switzerland}

\author{D.~V\'enos}
\affiliation{Nuclear Physics Institute, Academy of Sciences of the Czech Republic, CZ 250 68, \v{R}e\v{z} near Prague, Czech Republic}

\author{A.~Vollhardt}
\affiliation{Physik-Institut, Universit\"{a}t Z\"{u}rich, Z\"{u}rich, 8057, Switzerland}

\date{\today}% It is always \today, today, but any date may be explicitly specified

\begin{abstract}
A difficult task with many particle detectors focusing on interactions below $\sim$100\,keV is to perform a calibration in the appropriate energy range that adequately probes all regions of the detector.  Because detector response can vary greatly in various locations within the device, a spatially uniform calibration is important.  We present a new method for calibration of liquid xenon (LXe) detectors, using the short-lived $^{83\mathrm{m}}$Kr.  This source has transitions at 9.4 and 32.1\,keV, and as a noble gas like Xe, it disperses uniformly in all regions of the detector.  Even for low source activities, the existence of the two transitions provides a method of identifying the decays that is free of background.  We find that at decreasing energies, the LXe light yield increases, while the amount of electric field quenching is diminished.  Additionally, we show that if any long-lived radioactive backgrounds are introduced by this method, they will present less than 67$\times$10$^{-6}$\,events/kg/day in the next generation of LXe dark matter direct detection searches.
\end{abstract}

% PACS, the Physics and Astronomy Classification Scheme.

\maketitle
\newcommand{\Leff}{$\mathcal{L}_{\mathrm{eff}}$}
\newcommand{\Kr}{$^{83\mathrm{m}}$Kr~}
\newcommand{\Krns}{$^{83\mathrm{m}}$Kr} %no space after the Kr
\newcommand{\Co}{$^{57}$Co~}
\newcommand{\Cons}{$^{57}$Co}
\newcommand{\Rb}{$^{83}$Rb~}
\newcommand{\Rbns}{$^{83}$Rb}

\section{Introduction}
\label{sec:intro}

Liquid xenon (LXe) particle detectors are commonly used in the fields of dark matter direct detection\,\cite{Alner:2005pa,Alner:2007ja,Lebedenko:2008gb,Angle:2007uj,Angle:2008we}, neutrinoless double-beta decay searches\,\cite{leport-2007-578}, collider experiments\,\cite{Papa:2008zz}, and in nuclear medicine\,\cite{Giboni:2007zz,Lopes:1999bv}. In these applications, energy deposition is measured by counting scintillation photons, ionization, or both.  The high electron mobility\,\cite{Miller:1968zz} and scintillation properties\,\cite{Doke2002} of this liquid contribute to its versatility.  LXe response depends on the electronic stopping power ($dE/dx$) to the incoming particle, which itself depends on both the identity and energy of the particle.  It is therefore necessary to calibrate a detector's energy scale with a source whose response is known relative to the particles under study.  One common such ``reference source'' is \Cons, which emits $\gamma$-rays predominantly at 122\,keV.

Dark matter direct detection experiments search for low energy nuclear recoils caused by the scattering of weakly interacting massive particles (WIMPs) with atomic nuclei\,\cite{Bertone:2004pz,Jungman:1995df,Lewin199687}.  There are two main problems involved in using \Co to calibrate LXe detectors for this application.  The first is that the $\gamma$-ray energy is much higher than the recoiling nuclei energy produced by WIMP interactions.  Second, the attenuation length of 122\,keV $\gamma$-rays in LXe is $\sim$2.5\,mm, and hence the energy deposition will be highly localized as compared with the tens of cm typically characterizing the size of such detectors.  The two problems are actually compounded, because the attenuation length of $\gamma$-rays decreases as their energy decreases, and therefore sources providing lower-energy $\gamma$-rays will give an even more localized response than \Cons.  The topic of localization is an issue for point sources placed outside the detector, but also for point sources placed inside the detector.  In the latter case, the source must be attached to a mounting device of some kind; for low energy $\gamma$-ray sources, any device used for this purpose will likely block some of the scintillation light and potentially distort any existing applied electric fields.  It is therefore not possible to calibrate a detector with an internal point source under the same conditions that would exist during the actual measurement.  To avoid these difficulties, short-lived noble gas sources can be used which diffuse uniformly in LXe.  The XENON10 experiment used the metastable $^{131\mathrm{m}}$Xe\,\cite{Angle:2007uj,Ni:2007ih}.  This source solves the second problem (spatial localization), but its 164\,keV transition does not overcome the problem of an appropriate energy scale.  Additionally, due to its half-life of twelve days, the detector must sit for approximately 2.5 months following a calibration until the source activity has dropped to 1\% of its initial value.

A promising alternative solution is to use the metastable \Krns, first proposed in \cite{McKinsey_DUSEL}.  This source has been used for calibrations of detectors in the Large Electron-Positron Collider\,\cite{Decamp1990121,DeMin:1995tk}, as well as in the KATRIN experiment which attempts to measure the electron neutrino absolute mass\,\cite{Angrik:2005ep}.  \Kr should diffuse uniformly in a LXe detector, addressing the issue of spatial uniformity (while perfect uniformity of deposition is not explicitly required, a uniform deposition allows for calibration of the full detector in the minimum time).  Additionally, its two de-excitation lines at 9.4 and 32.1\,keV lie in the energy range of interest for dark matter direct searches, and its half-life of only 1.8~hours allows for a short turnaround time following measurement.  Simultaneous with the work presented here, another group has performed a demonstration of a similar \Kr introduction method in a single-phase LXe chamber\,\cite{Kastens:2009pa}.  Here we present a successful implementation of this calibration source in a small dual-phase LXe prototype detector.  Furthermore, we show results of measurements of the LXe energy scale linearity, evolution of energy resolution with energy, effects of LXe response under applied electric fields, and set limits on the level of long-lived radiocontaminants introduced by this method.

\section{Experimental Methods}
\label{sec:technique}
\subsection{The \Kr source}
\label{sec:technique:source}
\begin{figure}[htb]
  \begin{center}
   \includegraphics[width=0.5\textwidth]{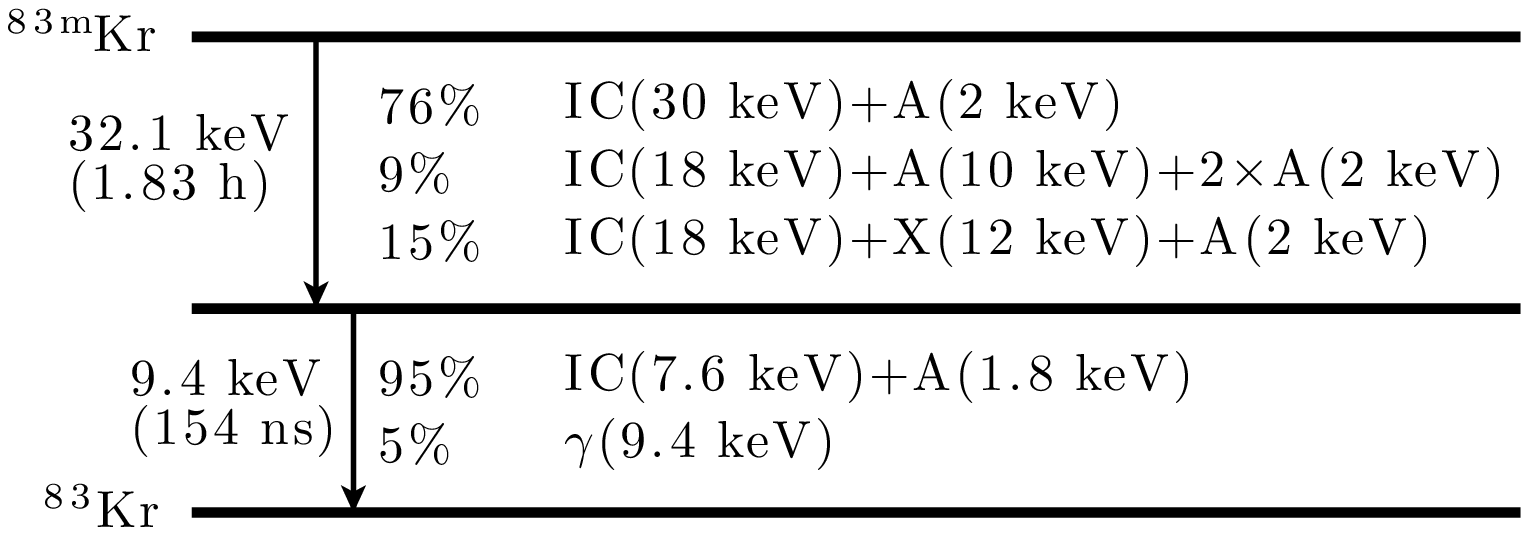}
   \caption[]{\label{Kr_decayScheme}The decay scheme and branching ratios of \Krns.  The decay always passes through two transitions, giving mostly internal conversion (IC) and Auger (A) electrons.  A small amount of the energy is carried by gamma-($\gamma$) and X-rays (X)\,\cite{DeMin:1995tk}.}
  \end{center}	
\end{figure}

\Kr is produced by the decay of \Rb via pure electron capture.  Following this process, the nucleus is left in any of $^{83}$Kr's many excited states lying below the $Q$-value of the Rb decay (910\,keV).  Regardless of the initial krypton excited state, the nucleus rapidly de-excites within picoseconds to the isomeric state \Krns, located 41.5\,keV above the ground state.  
Isomeric krypton decays with a half-life of 1.83\,h to the first $^{83}$Kr excited state (9.4\,keV), which then decays to the ground state with a half-life of 154\,ns\,\cite{Ekstroem}.  The decay scheme of \Kr is shown in Figure\,\ref{Kr_decayScheme}, indicating that most of the released energy is carried by internal conversion and Auger electrons\,\cite{DeMin:1995tk}.  The 6\,kBq \Rb source used in this study was produced at the Nuclear Physics Institute, \v{R}e\v{z} (Czech Republic).  This institute also provides \Rb for the KATRIN experiment\,\cite{Angrik:2005ep}.  
The parent \Rb is produced in the U-120M cyclotron from the reaction $^{\mathrm{nat}}$Kr(p,xn)$^{83}$Rb by irradiating a medium-pressure krypton target with 27\,MeV protons.  The product, deposited on the target chamber walls, is then washed into several tens of milliliters of high purity water ($<$0.07\,$\mu$S/cm).  An appropriate amount of the target eluate is then absorbed in zeolite beads (2\,mm diameter, Merck), which acts as a molecular sieve.  Zeolite was chosen due to its ability to allow for efficient emanation of \Kr in vacuum, while exhibiting high retention of the mother \Rb in its porous structure.  For details on the production process, we refer to~\cite{Venos2005323}.  In addition to \Rbns, $^{84}$Rb and $^{86}$Rb isotopes are also produced, however, they decay to stable Kr isotopes and hence introduce no radioactive backgrounds.  Since \Rb decays with a half-life of 86.2\,days, the source strength decreased to $\sim$3\,kBq by the end of these measurements.

\subsection{Detector System}
\label{sec:technique:detector}
\begin{figure}[h]
  \begin{center}
   \includegraphics[width=0.45\textwidth]{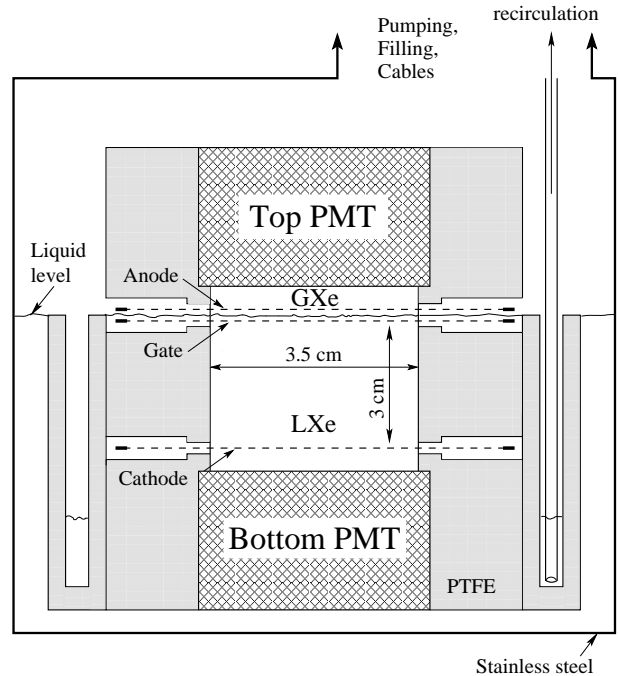}
   \caption[]{Schematic of the dual-phase X\"urich detector. The PTFE structure holds the PMTs and grid electrodes (see text), defining an active region 3.5\,cm in diameter and 3\,cm high.\label{SchemDet}}
  \end{center}	
\end{figure}

The `X\"{u}rich' detector, used for studying the \Kr decay is shown schematically in Figure \ref{SchemDet}.
It is a dual-phase (liquid and gas) LXe time projection chamber (TPC).  The system was built at the Universit\"{a}t Z\"{u}rich specifically for testing liquid xenon's response to low energy interactions.  The stainless steel vessel in Figure\,\ref{SchemDet} is located inside a vacuum cryostat, with cooling provided via a copper cold finger immersed in liquid nitrogen.  The temperature and pressure are held constant at 175\,K and 1.8\,bar, respectively, and the detector was operated stably for approximately 2.5 months.  X\"{u}rich's cylindrical active region, 3.5\,cm in diameter and 3\,cm in height (80.8\,g of LXe), is defined by a polytetrafluoroethylene (PTFE) cylinder on the perimeter and grid electrodes above (gate) and below (cathode).  A third grid electrode (anode) is located above the gate grid, with the liquid level lying between the gate and anode grids.  Two Hamamatsu R9869\,\cite{hamamatsu} photomultiplier tubes (PMTs) view the active volume, one from below and one from above.  A total of 1.76\,kg is used to fill the stainless steel vessel.

The cathode and gate grids apply an electric field of typically $\sim1$\,kV/cm which is used to drift electrons away from an interaction site towards the gate grid.  Once the electrons pass through the gate grid, they arrive at the liquid surface and are extracted to the gas by an electric field of $\sim10$\,kV/cm that then accelerates the electrons through the gas until they collect on the anode grid.  The high voltage applied to the grids is supplied by a CAEN model A1526 module.  During their transit through the gas, the electrons collide with Xe atoms with sufficient energy to produce scintillation light.  Therefore, the typical result of a particle interaction is a prompt scintillation signal (S1) emitted from the interaction site itself, followed by a delayed scintillation signal (S2) produced as the electrons travel through the gas under the influence of the extraction field.  In this way, both the scintillation and ionization signals are measured by the PMTs.  This technique is used for charge readout because it provides superior amplification over more traditional methods\,\cite{bolozdynyaDP:1999,yamashitaDP:2003}.  Additionally, the $z$-position of the event can be inferred by the delay time between the S1 and S2 signals since the electron drift velocity is well known as a function of applied field\,\cite{Miller:1968zz}.

\begin{figure}[h!]
  \begin{center}
   \includegraphics[width=0.45\textwidth]{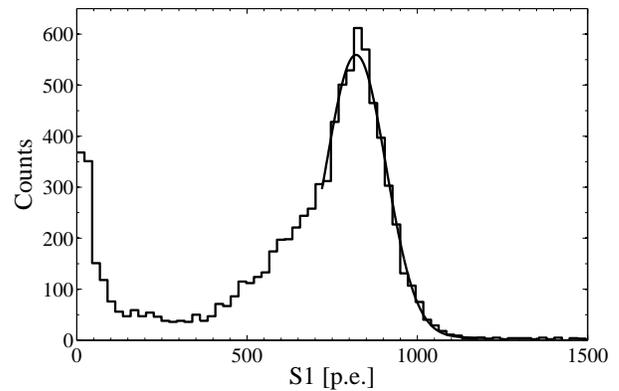}
   \caption[]{\label{co57_spect}Spectrum obtained from \Co at zero field and Gaussian fit.  The run-averaged light yield is 6.38\,p.e./keV.}
  \end{center}	
\end{figure}

The \emph{light yield} is defined as the number of photoelectrons (p.e.)~emitted from the PMT photocathodes per unit energy, and is customarily quoted based on the primary emission of \Cons.  When X\"urich is operated in single-phase mode with the liquid level above the top PMT, the \Co source produces $\sim$10\,p.e./keV.  In the dual-phase mode used in this study, where the liquid level lies below the top PMT, some of the scintillation light is refracted and reflected at the liquid surface.  Though some of the reflected photons may be detected by the bottom PMT, roughly 35\% are lost.  The result is a significantly larger S1 signal in the bottom PMT compared to the top (70\% on bottom, 30\% on top), and an overall reduced light yield as compared with the value taken in single-phase mode.  We measure the dual-phase \Co zero-field light yield to be $6.38\pm0.05(\mathrm{stat})\pm0.36(\mathrm{sys})$\,p.e./keV, with 11.5\% resolution ($\sigma/\mu$).  The systematic uncertainty is taken from the fluctuations of this light yield throughout the run, and the statistical uncertainty is the combination of fit uncertainties from each \Co zero-field data set.  The spectrum obtained from one \Co calibration is shown in Figure \ref{co57_spect}.

The PMTs were produced with photocathodes specifically designed to have their peak quantum efficiency ($\sim$25\% bottom, $\sim$35\% top) at 178\,nm, the wavelength of Xe scintillation light\,\cite{Jortner:1972}.  Negative high voltage of approximately $-$900\,V, supplied by a NHQ 225M high voltage module, is distributed to the photocathodes and 12 dynodes by voltage dividers made in-house on a PTFE substrate with a total impedance of $\sim$140\,M$\Omega$.  Each PMT is calibrated \emph{in situ} by a pulsed blue LED, periodically and before each measurement, to obtain the single p.e.~response, and hence the PMT gain ($\sim$4$\times$10$^6$ at these voltages).  The PMT signals are fed to an external fast voltage amplifier (Phillips 777) having two outputs; one output is connected directly to the ADC (Acqiris model DC436 100\,MS/s), while the other is fed to the triggering system.  The triggering system is set to discriminate each PMT individually at $\sim$1.5\,p.e., requiring coincidence between the two PMT channels.  The efficiency of this trigger, determined by Monte Carlo method, is 50\% at 1.6\,keV, rising to 95\% at 4\,keV and 99\% at 5.4\,keV, well below the 32.1\,keV events that provide the trigger in a \Kr decay.  In the offline signal processing, the pulse areas are converted to a value in number of p.e. by using the measured PMT gains.

\begin{figure}[htb]
  \begin{center}
   \includegraphics[width=0.45\textwidth]{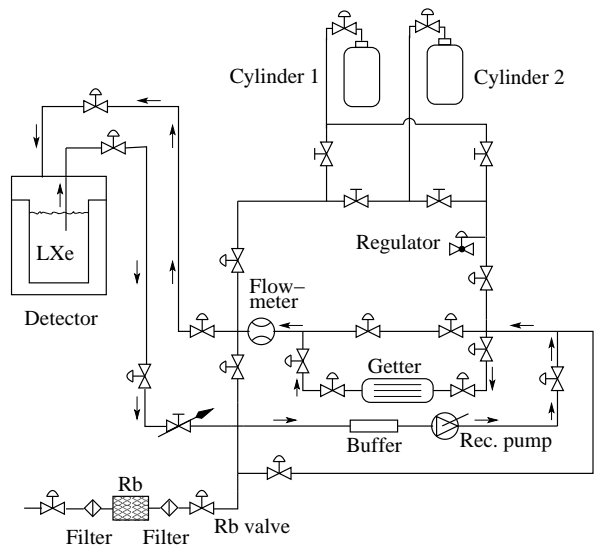}
   \caption[]{Schematic of the gas system. The recirculation pump draws liquid up through the recirculation siphon tube in the LXe chamber where it is vaporized while leaving the cryostat space, passed through a getter for purification, and the recondensed inside the chamber.  The zeolite beads containing \Rb are located inside the chamber labeled `Rb' and exposure to the gas system is controlled by the valve labeled `Rb valve'.\label{GasSystem}}
  \end{center}	
\end{figure}

In order to maintain the purity of LXe at a level sufficient for optimal light and charge collection, the xenon is constantly purified with a hot metal getter, SAES Mono Torr model PS3MT3R1.  The purification is handled by the gas system shown schematically in Figure\,\ref{GasSystem}. Liquid is drawn up from the detector by a closed diaphragm recirculation pump, Enomoto MX-808ST-S, where it is vaporized upon leaving the cryostat.  After passing through the pump, it passes through the getter and back into the X\"{u}rich detector where it is recondensed.  In this way, the xenon is constantly purified in a closed recirculation loop.

\Kr is introduced into the flow of the closed recirculation circuit by means of a single port with a valve.  The zeolite beads containing the \Rb reside in a small chamber filled with the same xenon gas that flows in the gas system.  Gaseous \Kr emanating from the \Rb decay may then diffuse into the recirculation circuit, its introduction being easily controlled by either opening or closing the valve at the port, denoted as the Rb valve.  Due to the rather long half-life of \Rb (86.2\,d), it is imperative that no trace of this mother radionuclide enters the system if it is to be used in a low background experiment.  Rb might potentially enter the system by one of two ways: as a vapor, which is very unlikely since its volatility under common laboratory temperatures even under vacuum is not significant; or as an aerosol formed from small particles of the zeolite itself.  Aerosol breakthrough is not entirely excluded, and therefore a 0.5\,$\mu$m aerosol filter is placed between the Rb chamber and the Rb valve in order to prevent any \Rb from entering the recirculation loop.  Measurements done to assess the level of \Rb introduced in the the system are discussed in sections \ref{sec:results} and \ref{sec:discussion}.

\section{Analysis and Results}
\label{sec:results}

Once the \Kr has entered the LXe, a 32.1\,keV transition might occur in the active region, which will then be followed by the 9.4\,keV transition.  A \Kr decay is, therefore, indicated by two S1 pulses whose separation in time is characterized by a decaying exponential with $t_{1/2}$=154\,ns.  Some of these transitions will occur too close in time to be resolved separately, giving a single 41.5\,keV pulse; however, the strength of this signal is well below the background level in the X\"{u}rich detector.
\begin{figure}[htb]
  \begin{center}
   \includegraphics[width=0.49\textwidth]{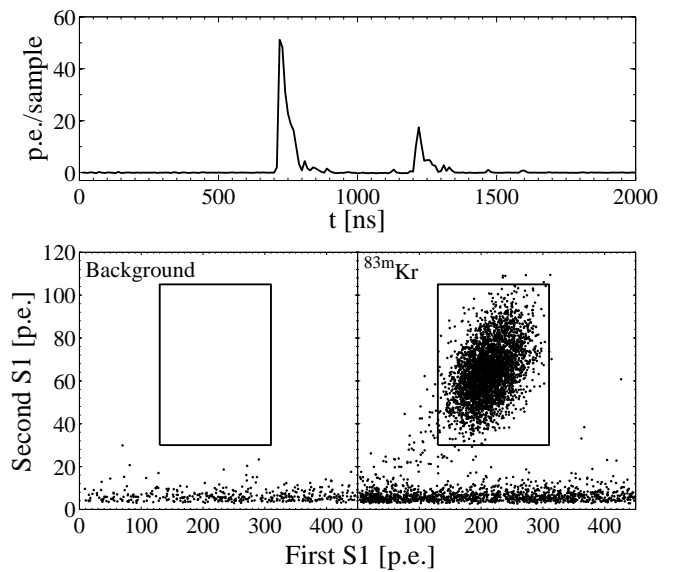} 
   \caption[]{\label{S1S1} (\textit{Top}) PMT output from a \Kr decay.  In this double pulse of primary scintillation light (S1), the first pulse corresponds to the 32.1\,keV transition with the second pulse resulting from the 9.4\,keV transition.  (\textit{Bottom}) The area of the first S1 pulse versus the area of the second, for events showing this characteristic two-pulse structure.  Shown are distributions taken before Rb exposure (`Background') and during Rb exposure (`\Krns'), demonstrating that the population of \Kr decays is clearly separated from background events.  The box represents the energy cuts used as the \Kr acceptance window.}
  \end{center}	
\end{figure}
On the other hand, many of the \Kr decays have a double S1 structure, while only a small fraction of non-\Kr decay events share this feature.  An example of the PMT response from a \Kr decay is seen in Figure \ref{S1S1} (\textit{top}).  

The events with such a double S1 structure are shown from one data set in Figure \ref{S1S1} (\textit{bottom}), with the area of the first pulse plotted versus the area of the second pulse.  In this space, it is evident that the \Kr decays form a population of events that is clearly separated from background.  The box indicates the energy cuts for first and second S1 pulses used to identify \Kr decays; before opening the Rb valve, background data show no events within this box.  After the Rb valve has been opened, the rate of \Kr decays in the total LXe volume increases to the 20\,Bq level in roughly 10\,h.  In order to further check that these are indeed \Kr decays, the distribution of S1 delay times (i.e.~the time between the first and second S1 pulses), $\Delta t_{\mathrm{S1}}$, of events within the box of Figure \ref{S1S1} (\textit{bottom}) is fit with a decaying exponential.  
\begin{figure}[htb]
  \begin{center}
   \includegraphics[width=0.48\textwidth]{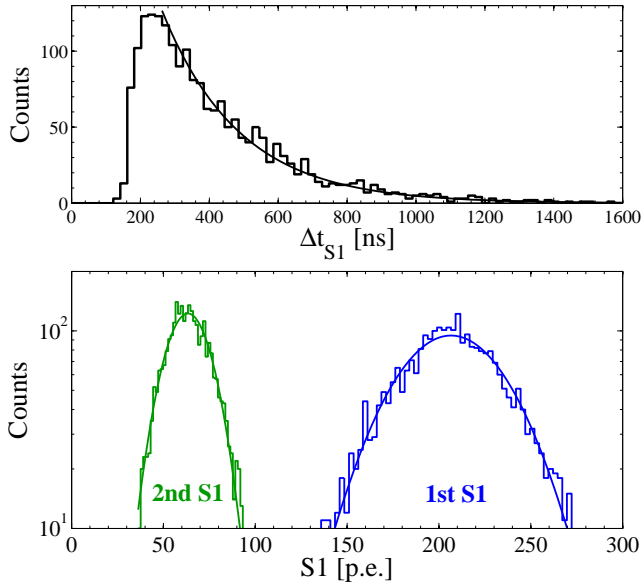} 
   \caption[]{\label{Kr_Spectra}(\textit{Top}) The distribution of delay times between first and second S1 pulses for events in the \Kr acceptance window.  An exponential fit to the distribution gives a half-life of $156\pm5$\,ns, consistent with the published value of 154\,ns.  (\textit{Bottom}, color online) Spectra from the two \Kr transitions, summed over all runs taken at zero field.}
  \end{center}	
\end{figure}
The result of the fit, shown in Figure \ref{Kr_Spectra} (\textit{top}), gives $t_{1/2} = 156\pm5$\,ns, consistent with the published value of 154.4$\pm$1.1\,ns\,\cite{Ekstroem}.  This excellent agreement validates the claim that these events are indeed caused by \Kr decays.  

Due to the shaping of the PMT signals by the various DAQ components, multiple S1 pulses that are delayed by less than $\sim$100\,ns cannot be separately resolved.  Additionally, the signal is required to be `clean' (i.e.~flat baseline) two samples before and after the pulse, in order to register as a positive S1 identification during the offline processing of the data.  This makes the efficiency for detecting multiple S1 pulses less than unity for $\Delta t_{\mathrm{S1}}<250$\,ns, as is obvious from Figure \ref{Kr_Spectra} (\textit{top}). Therefore, the double S1 selection cut detects \Kr decays with an efficiency of approximately 32\% under these conditions.

The spectra, in p.e., obtained at zero field from the two transitions of \Kr are displayed in Figure \ref{Kr_Spectra} (\textit{bottom}).  A Gaussian function is fit to each spectrum that is used to determine the light yield and energy resolution, shown in Table\,\ref{tab:LY_vs_En}.
\begin{table*}[ht!]
\begin{center}
\caption{\label{tab:LY_vs_En}The measured zero-field light yield (L.Y.), peak resolution (Res.), and field dependence fit parameters, $a_i$.  The row following 41.5\,keV gives the charge collection of the summed signal.  Uncertainties shown in light yield are statistical only; because these two peaks are taken from identical events, their systematic uncertainties are highly correlated, and hence do not affect the significance of the relative rise in light yield.}
	\begin{tabular}{ r@{.}l | c | c | c | c | c }
		\hline \hline 
		\multicolumn{2}{c}{$E$\,(keV)}\vline
		& \multicolumn{1}{c}{L.Y.(p.e./keV)}\vline
		& \multicolumn{1}{c}{Res.~($\sigma/\mu$)}\vline
		& \multicolumn{1}{c}{$a_1$}\vline
		& \multicolumn{1}{c}{$a_2$\,($10^{-4}$cm/V)}\vline
		& \multicolumn{1}{c}{$a_3$} \\
		\hline
		9&4  & 6.74$\pm$0.06 & 20.0\% & -0.35$\pm$0.06 & 6.3$\pm$3.0 & 1\\ %\hline
		32&1 & 6.43$\pm$0.04 & 14.4\% & -0.55$\pm$0.03 & 8.9$\pm$1.6 & 1\\ %\hline
		41&5 & --- & --- & 0.406$\pm$0.006 & 17$\pm$2 & 0.074$\pm$0.012\\
		123&6& 6.38$\pm$0.05 & 11.5\% & -0.679$\pm$0.007 & 12.6$\pm$0.5 & 1\\
		\hline \hline
	\end{tabular}
\end{center}
\end{table*}
As mentioned in section \ref{sec:intro}, \Co emits primarily 122\,keV $\gamma$-rays.  However, there is a small additional contribution from 136\,keV.  The two lines, however, are not distinguishable from one another due to the detector's energy resolution and instead give a single peak, whose weighted average energy is 123.6\,keV.  The measurements suggest a rise in the light yield at lower energies, consistent with behavior previously observed in LXe\,\cite{Yamashita2004692} and also in the XENON10 detector\,\cite{PSorensen}.  The peak resolutions ($\sigma/\mu$) are also shown at zero field.

\begin{figure}[htb]
  \begin{center}
   \includegraphics[width=0.48\textwidth]{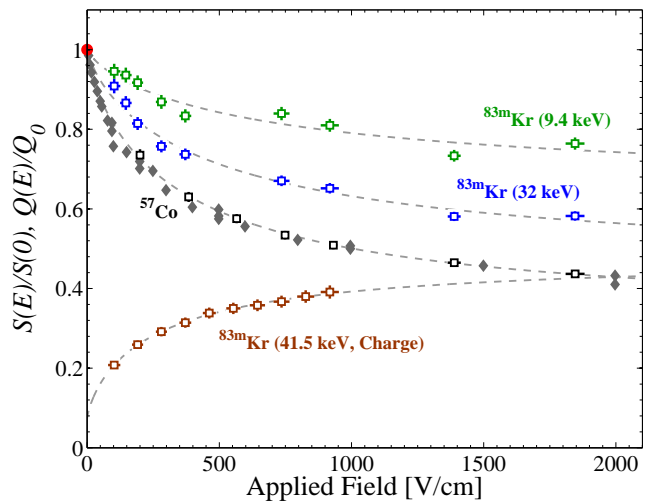} 
   \caption[]{\label{field_quench}(Color online) Field quenching, defined as the light yield of a spectral line divided by the light yield obtained at zero field, or $S(E)/S(0)$.  Data collected from \Co (open black squares) are consistent with those reported in \cite{Aprilenc:2006} (solid grey diamonds).  Dashed lines correspond to a fit parameterization described in the text.  Also shown is the field-dependent charge collection of the combination of both \Kr transitions, $Q(E)/Q_0$; the two transitions occur too close in time for their ionization signals to be individually resolved.}
  \end{center}	
\end{figure}

As mentioned in subsection \ref{sec:technique:detector}, most LXe detectors use an applied external electric field in order to collect electrons emitted from the interaction site.  As the applied field is increased, more and more electrons leave the interaction, suppressing the recombination process that contributes photons to the scintillation signal.  The result is that both the scintillation and ionization responses vary strongly with applied field, with the two signals exhibiting anti-correlation.  It is then crucial that the scintillation dependence on the applied field, called field quenching, be known quantitatively for any calibration sources.  Figure \ref{field_quench} shows the light yield as a function of the applied field, normalized to the zero field value, of the two \Kr transitions and the \Co line.  The uncertainty in the light yield is dominated by a 2\% systematic uncertainty taken from the measured fluctuations in the PMT gain over the duration of the run.  The horizontal positions are determined by electrostatic field simulations of the detector in each voltage configuration used; horizontal uncertainties are the 1-$\sigma$ variation of the field over the active volume.  The simulations were carried out using the COMSOL simulation package (commercially available)\,\cite{COMSOL}, and verified with software written in-house.

\begin{figure}[htb]
  \begin{center}
   \includegraphics[width=0.48\textwidth]{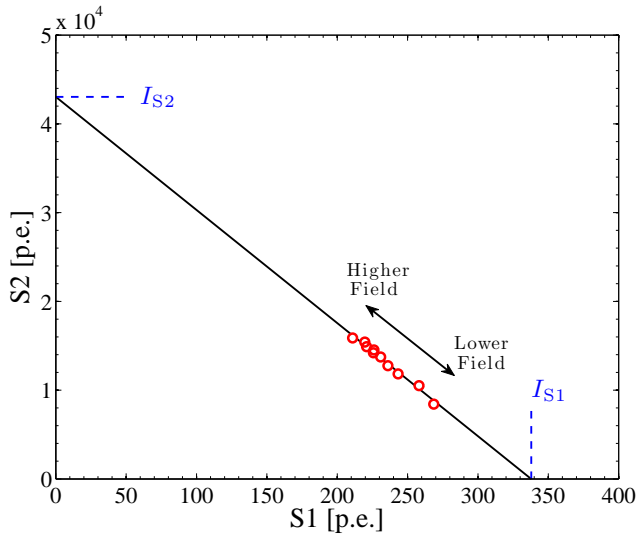} 
   \caption[]{\label{S2_v_S1}(Color online) The peak position in S2 versus S1 space for the 41.5\,keV emission of \Krns.  The data are taken from applied fields ranging from 100\,V/cm to 1\,kV/cm.  The line is fit to the data having vertical and horizontal intercepts $I_{\rm S2}$ and $I_{\rm S1}$, respectively; these intercepts indicate the location of $N_{ion}+N_{ex}$.}
  \end{center}	
\end{figure}

The time scale of the ionization signal, 1-2\,$\mu$s, does not permit the two \Kr transitions to be resolved separately, and instead the S2 signal contains the combination of charge emitted from both decays.  This 41.5\,keV summed-signal ionization yield is also shown in Figure \ref{field_quench} normalized to $Q_0$, the theoretical total amount of initial charge produced prior to electron-ion recombination.  This value is determined by plotting the S1 peak positions versus the S2 peak positions from data taken at various applied fields, shown in Figure \ref{S2_v_S1}.  As S1 and S2 are anti-correlated, these data lie along a line having negative slope, with the line's intercepts representing the total number of quanta, ions plus excitons ($N_{ion}+N_{ex}$).  For electronic recoils, the ratio of excitons to ions, $N_{ex}/N_{ion}$, is taken to be 0.06\,\cite{Takahashi.1771}, and hence $Q_0$ is 94.3\% the value of the S2 intercept.  The horizontal positions and error bars are determined in the same manner as those of the scintillation yield measurements, while the vertical error bars are instead dominated by the statistical errors in the peak fits and the uncertainty in $Q_0$.

The data are fit with a three-parameter function based on the Thomas-Imel box model for electron-ion recombination\,\cite{Thomas_j:1987}, given by,
\begin{equation}
\frac{S(E)}{S(0)},\frac{Q(E)}{Q_0} = a_1a_2E\ln\left(1+\frac{1}{a_2E}\right)+a_3 , 
\end{equation}
where $E$ is the electric field strength, and $S$, $Q$ are the scintillation and ionization yields, respectively.  This model is used only to provide a convenient parameterization of the data, and not to infer fundamental LXe physical properties from the results of the fits.  The $a_i$ are the parameters of the fit, shown in Table \ref{tab:LY_vs_En}.  Because the scintillation yields are normalized to the value at zero field, $a_3$ is unity and the function therefore contains only two free parameters for the scintillation yield data.  At decreasing energies, we observe a consistent decrease in the level of field quenching.

The energy of an event can also be measured by counting the total number of quanta, $N_{ion}+N_{ex}$.  This is called the combined energy scale (CES), and is constructed by forming a linear combination of the scintillation and ionization signals, $\alpha\mathrm{S1}+\beta\mathrm{S2}$, such that $n_{\gamma}=\alpha\mathrm{S1}$ and $n_{e}=\beta\mathrm{S2}$, where $n_{\gamma}$ and $n_{e}$ are the number of emitted photons and electrons, respectively.  The coefficients $\alpha$ and $\beta$ can be found from the plot of S1 versus S2 (Figure \ref{S2_v_S1}), by $\alpha=E/(WI_{\mathrm{S1}})$ and $\beta=E/(WI_{\mathrm{S2}}$), where $E$ is the deposited energy, $W$=13.5\,eV is the average energy required to produce a single quanta (electron or photon)\,\cite{Shutt:2006ed}, and $I_{\mathrm{S1(S2)}}$ is the S1(S2) intercept in units of p.e..  The CES has the advantage that it is not affected by correlated recombination fluctuations which dominate the S1 resolution over most energies\,\cite{Thomas_j:1987}, and hence gives an energy estimate with better resolution than S1 or S2 alone.  For example, the S1-only and S2-only peak resolutions of the 41.5\,keV peak taken at 500\,V/cm are 14.2\% and 20.1\%, respectively, while the resolution of the CES peak at this field is 10.0\%.

\begin{figure}[htb]
  \begin{center}
   \includegraphics[width=0.48\textwidth]{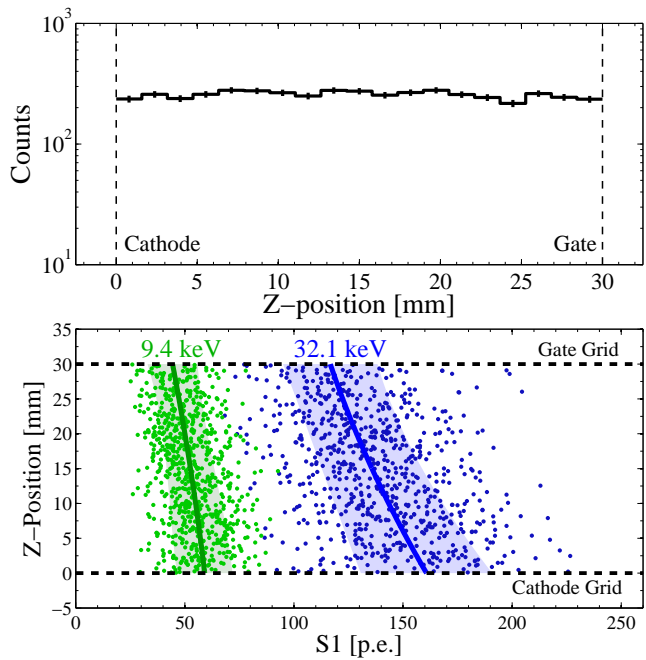} 
   \caption[]{\label{S1_vs_Z} (\textit{Top}) Rate of \Kr decays as a function of $z$-position, indicating a uniform concentration.  (\textit{Bottom}, color online) Measured $z$-dependence on the light yield from \Krns's two transitions taken at 1\,kV/cm.  The solid lines indicated the band centers, with $\pm1\sigma$ covered by the shaded areas.  Both lines show a light yield at the cathode that is a factor of 1.3 larger than at the gate grid.}
  \end{center}	
\end{figure}

The delay time between S1 and S2 gives the drift time of the electrons, and hence the $z$-position of the interaction.  One important motivation for using this source is that it should disperse uniformly in the active LXe volume, providing a spatially-uniform calibration.  The summed $z$-position distribution of \Kr events taken at drift fields from 100-1000\,V/cm is shown in Figure \ref{S1_vs_Z} (\textit{top}).  The observed $z$-dependent rate is flat with variations consistent with statistical fluctuations on each bin.  With this uniform calibration, the position-dependence of the detector's response can be measured and corrected for.  Most of the S1 signal is detected by the bottom PMT, and therefore one expects to see a light yield that is a monotonically decreasing function of $z$-position (i.e.~more light is collected from events occurring close to the bottom PMT than for events close to the top).  Figure \ref{S1_vs_Z} (\textit{bottom}) shows the light yield of the \Kr decays at all positions along the $z$-axis between the cathode and gate grids for the data run at 1\,kV/cm; solid lines are the band centroids, shaded bands cover $\pm1\sigma$.  In both transitions, the light yield at the cathode (bottom of active region) is a factor of 1.3 higher than the light yield at the gate grid (top of active region).  Data collected at lower applied fields shows consistency with this behavior.

Although the \Kr decays away in a matter of hours, the \Rb will live for nearly 1.5\,yr before decaying below 1\% of the initial activity.  If this technique is to be used in low-background experiments, it is then imperative that no \Rb atoms enter the system, and instead must remain trapped within the zeolite or the filter.  In order to test this, the valve to the \Rb chamber was closed.  The rate of \Kr decays is expected to decrease exponentially to zero during the following day; however, if \Rb has entered the system, the rate vs.~time will behave as an exponential decay with a vertical offset.  No such offset was observed in the \Kr rate following the closing of the Rb valve.  Indeed, 2.5\,h of data collected one day after closing the Rb valve resulted in zero observed events.  This null observation corresponds to a one-sided Poisson upper limit of $<$800\,$\mu$Bq (90\% C.L.) of residual \Kr in the active region.

\section{Discussion}
\label{sec:discussion}
Testing the low energy response of LXe is generally rather difficult, and therefore \Kr provides a unique tool for such measurements.  The rate of \Kr decays studied in this work is quite low as compared with the rate of background events due to natural radioactivity and cosmic rays, contributing roughly 0.5\% of the total triggers.  However, the double S1 structure of these decays, and energy cuts used, enable their measurement in a virtually background-free regime.  It is therefore not necessary to use a low background setup simply to study this weak source.  Simultaneous with this work, a demonstration of \Kr introduction to a single-phase LXe chamber by a similar technique has been performed by another group\,\cite{Kastens:2009pa}.

The light yield and energy resolution at low energies are of particular relevance for dark matter direct detection searches.  As indicated in Table\,\ref{tab:LY_vs_En}, the light yield increases at low energies.  Although an accurate quantitative understanding of this process is incomplete, the observed behavior can be understood qualitatively in the following manner.  The electronic stopping power of electrons in LXe increases at decreasing energies\,\cite{ESTAR}, and thus the ionization density produced by a recoiling electron increases along the track, with the highest densities concentrated at the track's end.  Because of this, the overall ionization density caused by a low energy recoiling electron will be greater than for an electron of higher energy.  The electrons and ions produced along the track will rapidly recombine and produce scintillation photons as they de-excite to their ground states.  The strength of recombination is correlated with the ionization density, because the characteristic electron-ion distance is shorter for higher ionization densities.  Even at zero applied electric field, not all of the electron-ion pairs produced by a recoiling particle will recombine to give scintillation photons\,\cite{Doke2002}.  It is then expected that the zero field recombination is stronger at lower energies (higher $dE/dx$), giving a higher overall light yield.

In \cite{Venos2005323}, 4\,h baking at 300\,$^{\circ}$C resulted in no observed loss of $^{83}$Rb.  The observation in this study of no \Kr decays, one day following the closing of the Rb valve, sets an upper limit of 800\,$\mu$Bq of residual \Kr \emph{inside the active region} of the X\"{u}rich detector.  Prior to this, the Rb valve had been opened for a total of 150 hours during the run.  The risk of Rb contamination increases with the amount of time that the valve is opened, and so this upper limit can be normalized to exposure time (150\,h).  Moreover, since the source is exposed to the gas system (and not the detector), the total activity in the LXe chamber should be independent of the detector size, and should instead depend on only flow rate and method of deployment.  The limit of 800\,$\mu$Bq in the active region (0.08\,kg) can be scaled to the total amount of LXe in the chamber (1.76\,kg), and normalized to the exposure, to give $<$120\,$\mu$Bq/h of residual \Kr in the whole liquid volume (assuming the \Kr concentration outside the active region is uniform and equal to the concentration inside the active region).  The branching ratio of $^{83}$Rb to $^{\rm 83m}$Kr is 75\%, which means this limit on residual $^{\rm 83m}$Kr is a limit of $<$160\,$\mu$Bq/h of residual $^{83}$Rb contamination.

To understand how this upper limit would affect an actual dark matter search, a 300\,kg detector with 100\,kg fiducial mass is taken as an example.  A detector of this size is typical of the proposed next generation of LXe dark matter searches\,\cite{aprileIDM:2008,gaitskellIDM:2008}.  An exposure to \Rb of 10\,h would be sufficient, under these conditions, to provide adequate statistics for such a calibration ($\sim$1000 Kr events/kg).  Our upper limit of \Rb contamination translates to a residual rate of $<$0.46\,decays/kg/day in this 300\,kg.  Even if this amount of \Rb was present in the system, the vast majority of decays would not introduce dangerous backgrounds.  In order for a background event to be `dangerous' (i.e.~appear in the WIMP signal acceptance window), it must have two features: (1) it must produce a single scatter event; (2) the event must deposit a small amount of energy that is within the WIMP search energy window.  An additional feature that dual-phase LXe TPCs like X\"{u}rich have is the ability to reject electronic recoils on an event-by-event basis at the level of $\sim$99.9\% based on the ratio S2/S1 \cite{Aprile:2010bt}.  However, statistical fluctuations can cause a small fraction of electronic recoil events to yield a S2/S1 ratio similar to values characteristic of a nuclear recoil from WIMPs, and thus the overall background level must be minimized as much as possible.  Any \Kr decays in the active volume would not present a problem because they would either have a double S1 structure (and could be vetoed on that basis), or would give 41.5\,keV, outside of the WIMP search region.  The only possibility for a dangerous background is from one of the $\gamma$-rays produced as the initial excited $^{83}$Kr decays to the metastable state.  These $\gamma$-rays are mostly emitted in the range of 500-600\,keV; again, to be dangerous they are required to single-scatter in the fiducial region, which is highly unlikely given their 3-4\,cm attenuation length.  With \Rb contamination at the level of our upper limit, Monte Carlo simulations indicate that 0.46\,decays/kg/day would contribute less than 67\,$\mu$DRU of single scatters in the WIMP search energy region (1\,DRU$\equiv$1\,event\,kg$^{-1}$\,day$^{-1}$\,keV$^{-1}$).  The projected $\gamma$ background rate in \cite{aprileIDM:2008} and \cite{gaitskellIDM:2008} due to natural radioactivity in the detector materials alone is roughly 1\,mDRU, fully fifteen times greater than our upper limit on the \Rb background.

After a calibration with \Krns, the stable $^{83}$Kr will remain in the system indefinitely unless some action is taken to specifically remove it.  However, the amount of Kr remaining from a 10\,h exposure as described above will be miniscule; less than 10$^6$ atoms total, which corresponds to a concentration of roughly 1 part in 10$^{21}$ for 300\,kg of Xe.  Even if this remaining concentration was higher, Kr will not adversely affect detector functions.  The transport of electrons through the Xe will not be diminished since Kr is chemically similar to Xe.  Additionally, Kr does not absorb Xe scintillation light\,\cite{baldini_absrb:1962} and therefore will not impede light collection.

In addition to \Rb contamination, water and oxygen trapped in the zeolite might also enter the system.  While these elements do not pose a problem in the context of radioactive backgrounds, they could affect the charge collection and light yield.  Before the Rb valve was initially opened, the Rb chamber was evacuated to the level of 10$^{-6}$\,mbar with a turbomolecular pump at room temperature.  The Rb valve was then open continuously for approximately four days, following which diminished charge collection was observed.  The Rb valve was then closed and the purification system allowed time to restore the LXe purity to a level adequate for negligible charge loss.  In subsequent measurements, the Rb valve was toggled in cycles of 20\,h open, 4\,h closed, with charge collection periodically monitored; no charge loss was measured under these conditions.  It is likely that the impurity content in the zeolite was depleted in the initial four days of exposure, and had left the system by the time the cycles of 20\,h exposure began.  In a subsequent run, the Rb chamber was baked for 24 hours at 120\,$^{\circ}$C prior to exposure, following which no effect on the electron lifetime was observed.  At no time was any effect on the light collection seen.

The X\"urich detector will be used in the future to study light and charge collection in LXe in response to nuclear recoils, scintillation pulse shape studies, and in studies related to research and development leading to a ton-scale LXe dark matter search.

\section{Summary}
\label{sec:summary}
This work presents a method to utilize the radioactive source \Kr in a LXe detector.  The two transitions of the \Kr decay produce two primary scintillation pulses, which allow these events to be clearly separated from background.  Additionally, the dual-phase detector used to study this source shows a spatially uniform distribution of decays, validating one of the main motivations for studying this source.  The data collected with this source show an increase in the light yield at decreasing energies, indicating a corresponding increasing level of zero-field electron-ion recombination.  A corresponding weaker field dependence of the light yield at decreasing energies is also observed.  An alternative energy scale based on the combination of scintillation and ionization signals was shown to provide superior energy resolution at 41.5\,keV compared with scintillation or ionization alone.

The technique used to introduce \Kr involves exposing the system to \Rbns, the presence of which in a low-background experiment could yield serious problems if contamination occurs.  This measurement provides an upper limit on the level of \Rb contamination introduced to the system of $<$160\,$\mu$Bq/h exposure.  In the context of the next generation of LXe dark matter direct detection searches, this upper limit translates to a maximum introduced background of 67\,$\mu$DRU, far below the 1\,mDRU of intrinsic backgrounds.

\begin{acknowledgments}
The authors wish to thank the machine shop at the Universit\"at Z\"urich for valuable help in fabrication of the X\"urich TPC, and the machine shop at the University of Florida for construction of the cryostat.  The work reported here was supported by the Universit\"at Z\"urich, by the Swiss National Foundation grant number 20-118119, and by the Volkswagen Foundation.  T.\,M.\,U.~acknowledges the Alexander von Humboldt Foundation for support through the Feodor Lynen scholarship.  Further support was provided by the Ministry of Education, Youth and Sports of the Czech Republic, under contracts LA318 and LC07050.
\end{acknowledgments}

\bibliography{xurich_kr83_arX}% Produces the bibliography via BibTeX.

\end{document}